\documentclass[10pt,fleqn,a4paper,twoside]{article}
\usepackage{amssymb}
\usepackage{amsmath}
\usepackage{abcm}
\def\shortauthor{T.E. Nazare, E.G. Nepomuceno and B.P.O Paiva} 
\def\shorttitle{On the Constructing Bifurcation Diagram of the Quadratic Map With Floating-Point Arithmetic}
\usepackage[utf8]{inputenc}
\graphicspath{{imagens/}}
\begin{document}
\fphead
\hspace*{-2.5mm}\begin{tabular}{||p{\textwidth}}
\begin{center}
\vspace{-4mm}
\title{COBEM-2017-XXIV\\
ON THE CONSTRUCTING BIFURCATION DIAGRAM OF THE QUADRATIC MAP WITH FLOATING-POINT ARITHMETIC
} 
\end{center}
\authors{Thalita Emanuelle de Nazar\'e} \\
\authors{Erivelton Geraldo Nepomuceno} \\
\authors{Bruno de Paula Ossalin Paiva} \\
\institution{Control and Modelling Group (GCOM), Department of Electrical Engineering, 
    Federal University of S\~ao Jo\~ao del-Rei, 
    S\~ao Jo\~ao del-Rei, MG, 36307-352, Brazil} \\
\institution{thalitanazare@gmail.com} \\
\institution{nepomuceno@ufsj.edu.br} \\
\institution{brunodepaula3p@yahoo.com.br} \\
\\

\abstract{\textbf{Abstract.} This paper presents an analysis on the effects of floating-point arithmetic on the constructing bifurcation diagram of the quadratic map. More precisely, we are interested in showing the dependence of initial conditions to obtain some specific features of the diagram. With this study, it was possible to observe that when there is a restriction regarding the initial condition, the results  present aspects with significant differences of the ones found in the literature regarding the behaviour of the map, consequently there is a considerable modification in its bifurcation diagram. We show that these difference are related to floating-point arithmetic}\\
\\
\\
\end{tabular}

\section{INTRODUCTION}

Dynamic systems have been investigated since the time of Newton, but the interest for numerical experiments grew considerably from the works realised by Lorenz \citep{Lorenz} in 1960. These works were carried out with the purpose of understanding the behaviour of nonlinear dynamic systems \citep{Grebogi}. Among the dynamic systems, those without an analytical solution are usually investigated by means of numerical computation, such as nonlinear systems with chaotic behaviour. Thus, the numerical computation plays a fundamental role in the analysis of nonlinear dynamic systems \citep{Lozi,Nepomuceno}
    
One of the most important way to study dynamic systems are the discrete maps.
These maps can also be seen as recursive functions, which allow the description and solution to a vast set of problems \citep{Feigenbaum}. They are the basis for the resolution of most nonlinear dynamic systems \citep{Lozi}. Using discrete maps it is possible to build bifurcation diagrams, which although much studied, do not have a well-defined set of rules for its application. As it has been already seen in \citep{Paiva}, the initial condition can be crucial in the constructing bifurcation diagram of the logistic map. In this paper, we are interested in doing a similar investigation of \citep{Paiva} for quadratic map \citep{Lorenz,Galias}, which has an important application in encryption of images based on chaos \citep{Ramadan,Kar}.

\section{PRELIMINARY CONCEPTS}
\label{sec:concepts}

\subsection{Recursive Functions}
Recursive functions that take the form $x_{n+1}=f(x_n)$ 
describe a wide range of problems. Being $\mathbb{I} \subseteq \mathbb{R}$ a metric space with $f:\mathbb{I} \rightarrow \mathbb{R}$ recursive function can be described as \citep{Nepomuceno}:

\begin{equation}
\centering
x_n=f(x_{n-1}). 
\label{eq01}
\end{equation}

\subsection{Fixed Point}

The fixed point $x^*$ of a map is the point such that $x_{n+1}=x_n=x^*$. That is, starting from $x^*$, 
remains in $x^*$ in the next iteration. Therefore, $x^*$ satisfies the following relation $x^*=f(x^*)$, \citep{Monteiro}.

\subsection{Quadratic Map}
 The quadratic map \citep{Lorenz} is described by the following function:
 \begin{equation}
  x_{n+1}=a-x_n^2, 
  \label{eq02}
\end{equation}

\noindent
where, \textit{n} 
is the number of iterations and \textit{a} is the bifurcation parameter.
\vspace{-3.3cm}
\begin{figure}[h!]
\centering
\includegraphics[angle=0, scale=0.5 ]{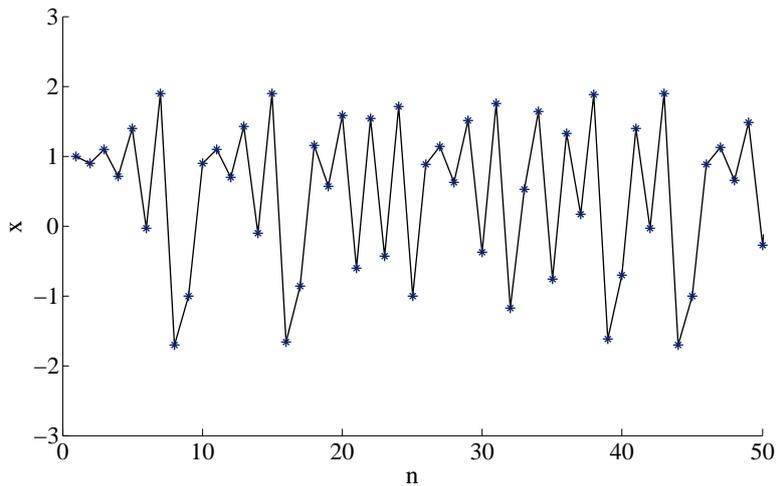}
\vspace{-3.3cm}
\caption{Result of the values of $x_n$ for n=50 and $a=1,9$.}
\label{fig1}
\end{figure}

\subsection{Bifurcation Diagram}

The term bifurcation is generally used to refer to the qualitative transition from regular to chaotic behavior by changing the control parameter \citep{Ramadan}, 
given any initial condition.The bifurcation diagram is used to study the system in function of its control parameter, allowing to know regions of the system that converge to bifurcation or even to chaos, depending on the parameter. \citep{Monteiro}.

\begin{figure}[h!]
\centering
\includegraphics[angle=0, scale=0.4 ]{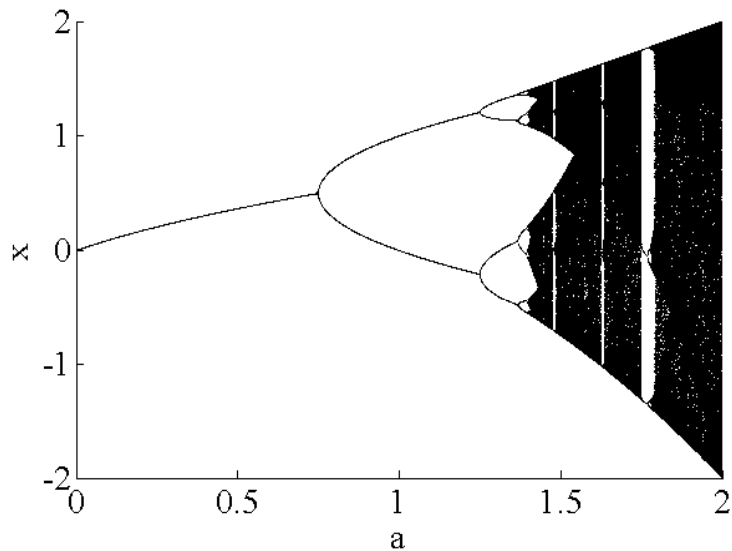}
\caption{Diagram of bifurcation for the quadratic map.}
\label{fig2}
\end{figure}

\subsection{Floating-Point Arithmetic}

Numerical computing is a critical part of dynamic systems analysis. Therefore, it is necessary to emphasize the IEEE 754-2008 standard \citep{IEEE} for floating-point arithmetic, since a large majority of numerical computation makes use of this norm. It establishes the rounding behavior, error handling and degree of accuracy of the calculations performed, causing the results to approach the expected value. Even with satisfactory results, numerical computation is not yet able to present results that are totally in agreement with reality, due to the fact that it has a finite representation of real numbers. \citep{Rodrigues,IEEE}.

\section{METHODOLOGY}
\label{sec:methodology}

The behaviour of the quadratic map is determined by the variation of the bifurcation parameter $a$, 
where $-2.5 < a <2.5$. Thus, given an initial condition belonging to the defined interval, the function will have its behaviour as shown in the bifurcation diagram of Fig. (1a).

Given the equation $x_{n+1}=a-x_n^2$ which describes the Quadratic Map, it is possible to obtain its fixed point  $x^*$ doing  $x_{n+1}=x_n=x^*$. Thus the map equation takes the following form:
 \begin{equation}
 \centering
  x^2 + x - a =0. 
  \label{eq03}
\end{equation}
From Eq.(\ref{eq03}), it is possible to  find a value for \textit{x} depending on the parameter \textit{a} which is a fixed point, shown by Eq. (\ref{eq04}).
\begin{equation}
\centering
    x_1=\frac{\sqrt{4a+1}-1}{2};
    x_2=-\left ( \frac{1+\sqrt{4a+1}}{2} \right )
    \label{eq04}.
\end{equation}

The use of the inverse of the function \eqref{eq03}, presented in Eq. (\ref{eq05}), allows to  verify the existence of a set of points that converge to the fixed point $x^*$. 

\begin{equation}
\centering
   f^{-1}(x^*)=\pm \sqrt{a-x^*}.
\label{eq05}
\end{equation}

\thispagestyle{empty} 
From Eq. (\ref{eq05}) the following result was obtained:
\begin{equation}
\centering
    x = \pm \sqrt{\left ( a -\frac{\sqrt{4a+1}-1}{2} \right )}.
    \label{eq06}
\end{equation}
Eq. (\ref{eq06}) is so used as the initial condition $x_0$ to build the bifurcation diagram. Thereby, 
it was possible to show that $\pm \sqrt{\left ( a -\frac{\sqrt{4a+1}-1}{2} \right )}$ belongs to the set of points that converge to the fixed point $x^*$.

\begin{equation*}
  x_1 = f(x) = a - \left [\sqrt{\left ( a -\frac{\sqrt{4a+1}-1}{2} \right )}\right ]^2 = \frac{\sqrt{4a+1}-1}{2};
\end{equation*}

\begin{equation}
\centering
  x_2 = f(x_1) = a - \left [\frac{\sqrt{4a+1}-1}{2}\right ]^2 = a - \left [\frac{4a+1 -2\sqrt{4a+1} +1}{4}\right] = \frac{\sqrt{4a+1}-1}{2};
 \label{eq07}
\end{equation}
\vspace{0.1cm}

\noindent
and $x_1 = x_2 = ... = x_n$, then $d(f^p(x),x^*) \to 0$ when $p \to \infty$.

\section{RESULTS}

\begin{figure}[h!]
\centering
\vspace{-3.3cm}
\includegraphics[angle=0, scale=0.5 ]{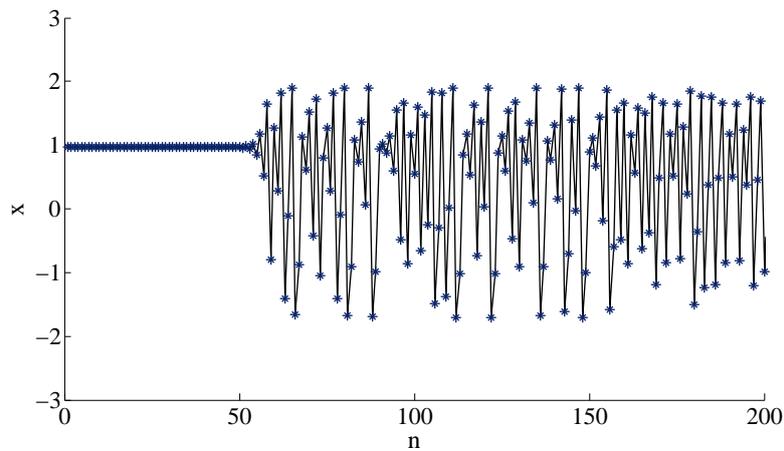}
\vspace{-3.3cm}
\caption{Result of the values of \textit{x} for n=200 and $a=1,9$.}
\label{fig3}
\end{figure}

The bifurcation diagrams are shown in Fig. (\ref{fig5}) analysis of the behavior of the map given a value for \textit{a}. When $a=1,9$ the map assumes an unstable fixed-point region, shown in Fig.(\ref{fig1}). The effect of the application of $ x_0=\sqrt{\left ( a -\frac{\sqrt{4a+1}-1}{2} \right )}$ as initial condition can be observed by Fig. (\ref{fig3}) where to $a=1,9$, constraint to the initial condition, it was possible to find a convergence for $x^*$, showing a result that is considerably different from that found in Fig. (\ref{fig1}), 
where no restrictions were made for the initial condition and only 50 iterations were performed.

\newpage

\begin{figure}[h!]
\centering
\includegraphics[angle=0, scale=0.4 ]{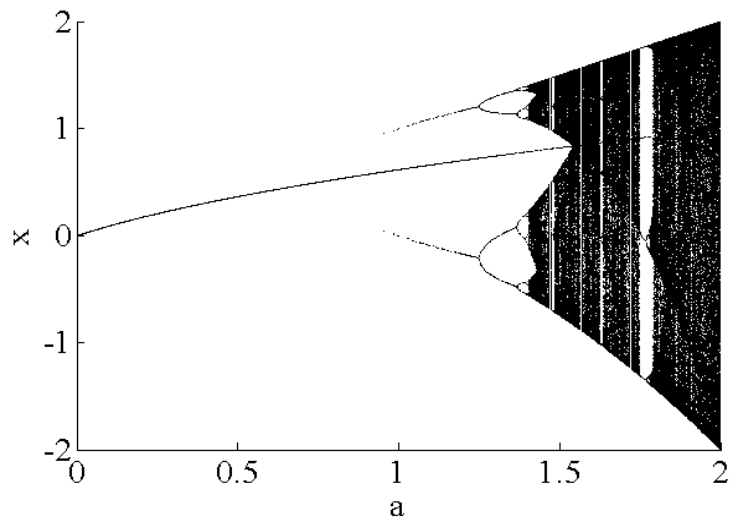}
\caption{Bifurcation Diagram for Initial Condition $x_0=\sqrt{\left ( a -\frac{\sqrt{4a+1}-1}{2} \right )}$.}
\label{fig4}
\end{figure}

\begin{figure}[h!]
\centering
\includegraphics[angle=0, scale=0.4 ]{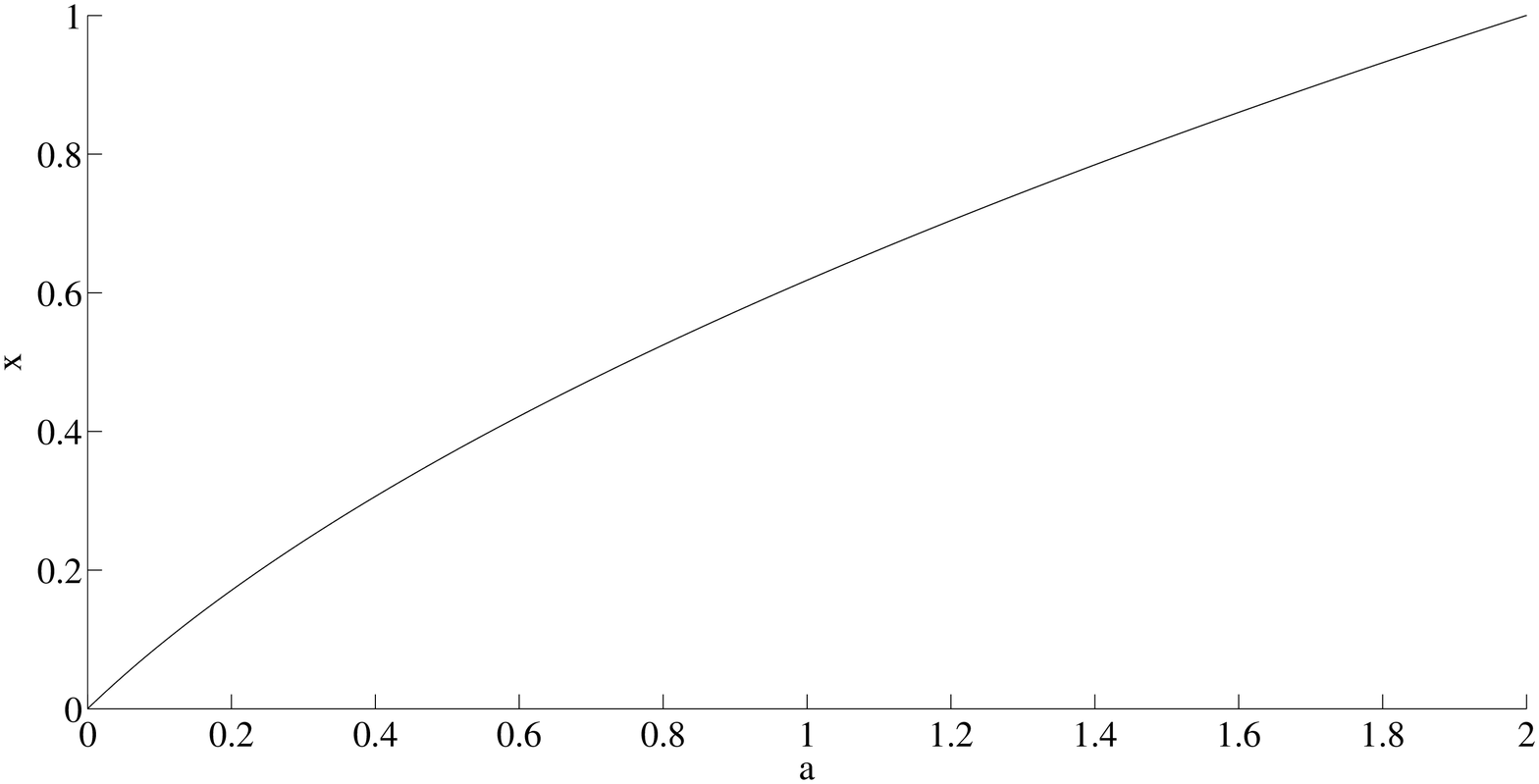}
\caption{Bifurcation diagram for initial condition $x_0=\sqrt{\left ( a -\frac{\sqrt{4a+1}-1}{2} \right )}$.}

\label{fig5}
\end{figure}

From these results, it is necessary to obtain a new bifurcation diagram from the initial condition $ x_0=\sqrt{\left ( a -\frac{\sqrt{4a+1}-1}{2} \right )}$. 
This diagram is shown in Fig.(\ref{fig4}). The diagrams shown in Figs.(\ref{fig4}) and (\ref{fig2}) were obtained in conventional manner in the literature using Matlab and double precision floating point arithmetic.

The result shown in Fig.(\ref{fig4}) is largely incorrect, although, due to its similarity with Fig.(\ref{fig2}), could be easily understood as correct.  With initial condition given by Eq.(\ref{eq06}), the correct answer is shown in Fig. (\ref{fig5}), which the transient is not discharged, and on the contrary, the bifurcation diagram has been built using only the first result of the iteration process, that is, using only $x_1$.  Similar results were found by \citep{Paiva}, for the logistic map.
\newpage

\section{CONCLUSIONS} 
\label{sec:conclusions}
After all the analyses, it is concluded that for the construction of the bifurcation diagram of the quadratic map the value of the initial condition has a fundamental role. The diagrams shown in Figs. (\ref{fig2}), (\ref{fig4}) and (\ref{fig5}) show different results. For Fig. (\ref{fig2}), the diagram was obtained following the methodology proposed in the literature for an equal initial condition $x_0 = 0.2$. While in the simulations of the diagrams shown in Figs. (\ref{fig4}) and (\ref{fig5}), a constraint associated with the parameter $ a $ has been applied. In Fig. (\ref{fig4} the numerical simulation did not present the expected results because using the proposed initial condition, for any value of $ a $, convergence occurs at the first iteration. However, it was observed that after a number of iterations, the result diverges. This divergence can be understood as the result of a numerical error due to the finite-precision of the floating-point arithmetic. Therefore,  it is necessary more care on the constructing bifurcation diagram of the quadratic map.

\section{ACKNOWLEDGEMENTS}

We thank CAPES, Fapemig and UFSJ for the support.
\section{REFERENCES} 
\bibliographystyle{abcm}
\renewcommand{\refname}{}
\bibliography{bibfile}

\section{RESPONSIBILITY NOTICE}

The authors are the only responsible for the printed material included in this paper.

\end{document}